\pdfoutput=1
\documentclass{article}

\usepackage{microtype}
\usepackage{graphicx}
\usepackage{subcaption}
\usepackage{booktabs}
\usepackage{array}
\usepackage{multirow}
\usepackage{colortbl}
\usepackage{enumitem}
\usepackage{hyperref}
\usepackage{xurl}

\usepackage[preprint]{icml2026}

\usepackage{amsmath}
\usepackage{amssymb}
\usepackage{mathtools}
\usepackage{amsthm}
\usepackage{float}
\usepackage[capitalize,noabbrev]{cleveref}

\newcommand{\method}{Speculative Listening}
\newcommand{\ours}{\method{}}
\newcommand{\best}[1]{\textbf{#1}}
\newcommand{\na}{--}
\newcommand{\x}{$\times$}

\icmltitlerunning{Decode Before You Hear: Speculative Listening for Full-Duplex Speech Language Models}

\begin{document}

\twocolumn[
\icmltitle{Decode Before You Hear: Speculative Listening for\\Chunk-Interleaved Full-Duplex Speech Language Models}

\icmlsetsymbol{equal}{*}

\begin{icmlauthorlist}
\icmlauthor{Yutong Bian}{cuhksz,amphion}
\icmlauthor{Yanglin Zhang}{amphion}
\icmlauthor{Yuxiang Wang}{cuhksz}
\icmlauthor{Luoyuan Zhang}{cuhksz}
\icmlauthor{Rongshen He}{cuhksz}
\icmlauthor{Zihao Wang}{cuhksz}
\icmlauthor{Junwen Qiu}{cuhksz}
\icmlauthor{Jiaqi Li}{cuhksz,amphion}
\icmlauthor{Zhizheng Wu}{cuhksz,amphion}
\end{icmlauthorlist}

\icmlaffiliation{cuhksz}{The Chinese University of Hong Kong, Shenzhen, China}
\icmlaffiliation{amphion}{Amphion Technology Co., Ltd}
\icmlcorrespondingauthor{Yutong Bian}{yutongbian02@gmail.com}
\icmlcorrespondingauthor{Zhizheng Wu}{wuzhizheng@cuhk.edu.cn}

\icmlkeywords{full-duplex speech language models, speculative decoding, speculative listening, real-time inference}

\vskip 0.3in
]

\printAffiliationsAndNotice{}

\begin{abstract}
Chunk-interleaved full-duplex speech language models listen and speak in fixed time chunks, and each chunk must be processed before the next one arrives. Their workload is asymmetric across chunks: listen chunks leave compute idle while awaiting audio, and speak and tool-call chunks require serial generation, which delays responses and tool calls. Using the idle compute for the next chunk requires decoding it before its audio arrives, under an assumed input. Taking silence as the assumed input and comparing the output with the output under real audio on four models, we observe two properties. First, the next audio chunk does not change the output in 95.3\% of chunks. Second, 88\% of changes begin at the first output token. Building on these properties, we propose \method{}, a training-free, lossless framework for input-time self-speculative decoding. During idle time, the model drafts the next chunk under silence and, for changes at the first token, a second branch from the runner-up first token. It keeps all drafts in a cross-chunk hypothesis pool and verifies them when the audio arrives. On our eight-task benchmark TickTock-Bench, \method{} speeds up speak chunks by 2.6--2.8\x{} and tool-call chunks by up to 23.4\x{}. On average, it reduces the decoding delay before a response starts by 73\%, to 74--241\,ms, close to the 200\,ms gap between human turns, and the maximum lag behind real time by 94.4\%.
\end{abstract}

\section{Introduction}
\label{sec:intro}
\begin{figure}[t]
  \centering
  \includegraphics[width=\columnwidth]{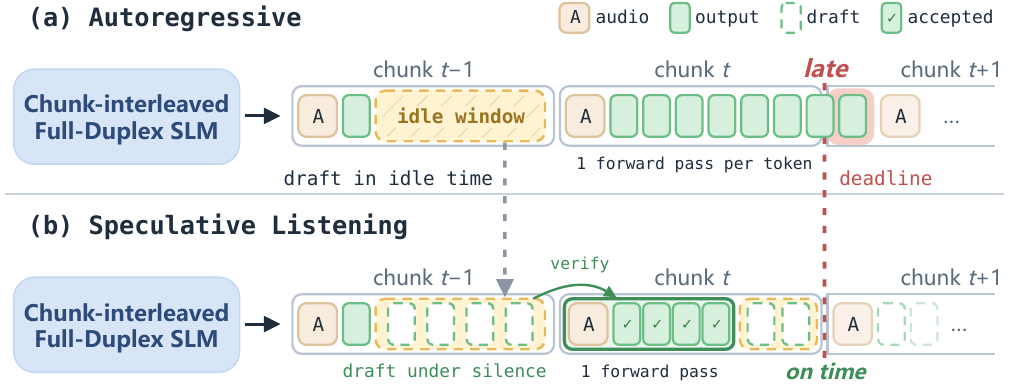}
  \caption{Decoding in a chunk-interleaved full-duplex speech language model. (a) Autoregressive decoding leaves compute idle while it waits for audio, and a long output then misses the chunk deadline. (b) \method{} drafts the next chunk under silence during the idle window and verifies the whole draft in one forward pass when the audio arrives.}
  \label{fig:overview}
\end{figure}

Full-duplex speech language models listen and speak at the same time~\citep{moshi,minicpmo45,tml_interaction}. A widely adopted design is the chunk-interleaved architecture~\citep{tdm,minicpmo45,tml_interaction,gander,venus,lws}, which divides a streaming conversation into fixed time chunks (\cref{fig:overview}). In each chunk, the model processes the newly arrived audio and decodes its output, which may keep listening, speak, or call a tool~\citep{gander,venus}. The audio of the next chunk arrives in real time, and the chunk duration acts as a hard deadline~\citep{tml_interaction}: a chunk that takes longer to decode delays the following chunks, and the conversation lags behind real time. The model then processes the user's speech late, and its responses become stale in the ongoing interaction.

Profiling four chunk-interleaved models chunk by chunk (\cref{fig:observations}a), we find their workload asymmetric across chunks. Listen chunks only process the newly arrived audio and leave compute idle for 70--89\% of the chunk duration while awaiting the next audio chunk (\cref{fig:overview}a). We call this unused time the \emph{idle window}. Speak and tool-call chunks require serial generation, and its length grows with the output: a long output, such as a tool call, takes tens of serial forward passes (\cref{fig:analysis}a), each bound by the memory bandwidth of loading model weights and the KV cache~\citep{shazeer2019}. Tool-call chunks miss the deadline, and a model with a written channel decoded alongside speech~\citep{lws} misses it on 95\% of speak chunks. Decoding also delays the start of each response: under AR, it adds 417--1089\,ms on average. Human speakers respond about 200\,ms after the other speaker ends~\citep{levinson2015}.

Prior work addresses neighbouring parts of this problem. Lossless speculative decoding~\citep{leviathan2023,chen2023} accelerates serial generation with draft models~\citep{eagle,medusa} or context lookup~\citep{pld,tokenrecycling} once the input is available. Recent speech systems compute during listening windows by training the model to emit reasoning tokens~\citep{chronothink} or by pre-generating candidate responses in turn-based cascaded pipelines~\citep{predgen}. In end-to-end chunk-interleaved models, the idle window has not yet been used to decode the output of the next chunk.

Using the idle window requires decoding a chunk before its audio arrives, that is, under an assumed input. We take silence as the simplest assumption, which we call the \emph{silence hypothesis}, and at every chunk across four models decode the output twice: once with the real audio and once with the audio replaced by silence (\cref{sec:acceptance}). Comparing the two outputs, we observe two properties (\cref{fig:observations}b,c):
\begin{enumerate}[leftmargin=*,itemsep=2pt,topsep=2pt]
  \item \textbf{The next audio chunk usually does not change the output.} The output under silence matches the real-audio output token for token in 95.3\% of chunks pooled over the four models.
  \item \textbf{When the output changes, it diverges at the first token.} Pooled over the four models, 88\% of divergences occur at the first output token, where most of the models decide to stay silent or start responding.
\end{enumerate}

\begin{figure*}[t]
  \centering
  \includegraphics[width=\textwidth]{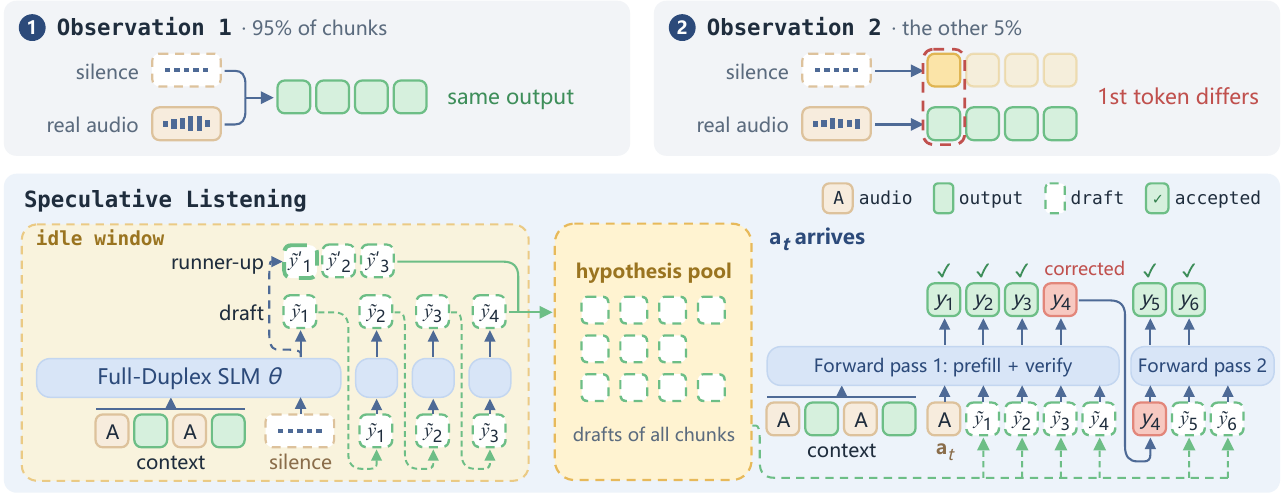}
  \caption{Overview of \method{}. Observation~1 motivates drafting under silence and verifying the draft in the audio prefill pass. Observation~2 motivates a runner-up branch at the first token and a hypothesis pool for later forward passes to continue from.}
  \label{fig:method}
\end{figure*}

Building on these properties, we propose \emph{\method{}}, a training-free and lossless framework for \emph{input-time self-speculative decoding} (\cref{fig:overview}b and \cref{fig:method}). During the idle window, the target model drafts the upcoming chunk under the silence hypothesis and keeps the drafts of all earlier chunks in a \emph{cross-chunk hypothesis pool}~\citep{lookahead}. A second draft branch, started from the runner-up first token, covers divergences at the first token. When the real audio arrives, one forward pass processes it and verifies the whole draft. If the output departs from the draft, later forward passes continue from a matching draft in the pool, such as the runner-up branch. Post-audio computation then grows only with changed tokens (\cref{sec:sl}).

We also introduce \emph{TickTock-Bench}, a Spec-Bench-inspired~\citep{specbench} unified benchmark and replay harness for full-duplex decoding under chunk deadlines. It covers eight reasoning, conversation, and agentic tasks (\cref{sec:bench}). Experiments on four models show that \method{} speeds up speak chunks by 2.6--2.8\x{} on average and tool-call chunks by up to 23.4\x{} (\cref{fig:case_long}b), reduces the maximum lag behind real time by 94.4\% on average (\cref{tab:backlog}), and shortens the decoding delay before a response starts by 73\% on average, to 74--241\,ms (\cref{sec:case}).

Our contributions are three-fold:
\begin{itemize}[leftmargin=*,itemsep=1pt,topsep=2pt]
  \item We identify two properties of chunk outputs in chunk-interleaved full-duplex models: the next audio chunk usually does not change the output, and changes concentrate at the first output token.
  \item We propose \method{}, a training-free and lossless decoding framework whose post-audio computation grows only with changed tokens, and which stays faster than autoregressive decoding under stochastic sampling and multi-session serving.
  \item We introduce TickTock-Bench, a benchmark and replay harness for decoding in chunk-interleaved full-duplex models, which enables a comparison of decoding methods under the same chunk deadlines.
\end{itemize}

\section{Related Work}
\label{sec:related}
\paragraph{Full-duplex speech language models.}
They listen and speak at the same time. Moshi~\citep{moshi} modeled the user's speech and its own as parallel streams. Chunk-interleaved models arranged input and output in one causal sequence: \citet{tdm} split the dialogue into time slices, OmniFlatten~\citep{omniflatten} flattened the streams into a single sequence, and MiniCPM-o 4.5~\citep{minicpmo45} and the interaction model of \citet{tml_interaction} processed streaming dialogue in fixed-duration chunks. Gander~\citep{gander} and Realtime-Venus~\citep{venus} added tool calls and external delegation to the stream, and LWS~\citep{lws} added a written channel that can be decoded in the same chunk as speech.

\paragraph{Speculative decoding.}
Speculative decoding~\citep{leviathan2023,chen2023} drafted several tokens cheaply and verified them in one forward pass of the target model, which preserves its output distribution. Drafts came from a smaller model, from heads trained on the target's hidden states as in Medusa~\citep{medusa} and EAGLE~\citep{eagle}, or from a subset of the target model's own layers, as in Draft \& Verify~\citep{draftverify} and Kangaroo~\citep{kangaroo}. Training-free methods drafted without extra parameters: prompt lookup decoding (PLD)~\citep{pld} matched $n$-grams in the context, Token Recycling~\citep{tokenrecycling} reused candidate tokens from earlier decoding steps, and lookahead decoding~\citep{lookahead} collected $n$-grams from Jacobi iterations in a pool. Spec-Bench~\citep{specbench} compared these methods under a unified protocol for text generation. In spoken dialogue, speculative generation was used to start responses earlier: RelayS2S~\citep{relays2s} let a fast duplex model draft the opening of a response that a cascaded pipeline can continue, and SpecSpeak~\citep{specspeak} started speech synthesis from draft tokens while the target model verified them. These methods draft once the input is available; \method{} is a training-free self-speculative method that drafts the output of a chunk before its audio arrives.

\paragraph{Computing ahead of the input.}
Human speakers plan their response while still listening, which keeps the gap between turns near 200\,ms~\citep{levinson2015}. Speech systems have used the listening time in several ways. Chronological Thinking~\citep{chronothink} trained a full-duplex model to emit reasoning tokens while listening. PredGen~\citep{predgen} pre-generated candidate responses from partial user input in a cascaded pipeline. Saguaro~\citep{saguaro} prepared speculations for likely verification outcomes while the current verification was still running. These methods compute ahead of their input with extra training, a cascaded pipeline, or a separate draft model; \method{} drafts the next chunk with the unchanged model and verifies it against the real audio.

\section{Method}
\label{sec:method}
In this section, we formalize decoding in a chunk-interleaved model, report two observations on chunk outputs, and present \method{}.

\subsection{Preliminary: Chunk-Interleaved Decoding}
\label{sec:prelim}
A chunk-interleaved model divides a conversation into time chunks $t = 1, 2, \dots$ of fixed duration. Let $\mathbf{a}_t$ denote the input tokens of chunk $t$, namely the encoded audio of the chunk together with any concurrent text input, and let $\mathbf{y}_t$ denote its output tokens. The model processes all chunks as one causal sequence, and the context of chunk $t$ is $\mathbf{x}_{<t} = (\mathbf{a}_1, \mathbf{y}_1, \dots, \mathbf{a}_{t-1}, \mathbf{y}_{t-1})$. Within a chunk, we drop the chunk index and write $\mathbf{y}_t = (y_1, \dots, y_{|\mathbf{y}_t|})$ with $y_i \in \mathcal{V}$. In most models, the first token $y_1$ decides whether the model keeps listening or starts an output, through an explicit listen/speak token in some models~\citep{minicpmo45,gander} and an end-of-chunk token in others; the remaining tokens carry the content, such as text for a speech decoder or a structured tool call. Under greedy decoding, a model with parameters $\theta$ generates
\begin{equation}
y_i = \arg\max_{y \in \mathcal{V}} \, p_\theta\left(y \mid \mathbf{x}_{<t}, \mathbf{a}_t, \mathbf{y}_{<i}\right).
\label{eq:ar}
\end{equation}
A forward pass of the model processes a block of input tokens and yields the distribution of the next token at each position. Autoregressive decoding prefills $\mathbf{a}_t$ and then decodes the output token by token. Each forward pass takes the token decoded by the previous one as input, and the forward passes run one after another. Their number after $\mathbf{a}_t$ arrives grows linearly with $|\mathbf{y}_t|$. A forward pass over a few tokens takes about as long as a forward pass over one token~\citep{shazeer2019,kim2023bild}. We call the time from the arrival of $\mathbf{a}_t$ to the end of the chunk's computation its \emph{post-audio latency}. A chunk whose post-audio latency exceeds the chunk duration misses its \emph{deadline} and delays the following chunks.

\begin{figure*}[t]
  \centering
  \includegraphics[width=\textwidth]{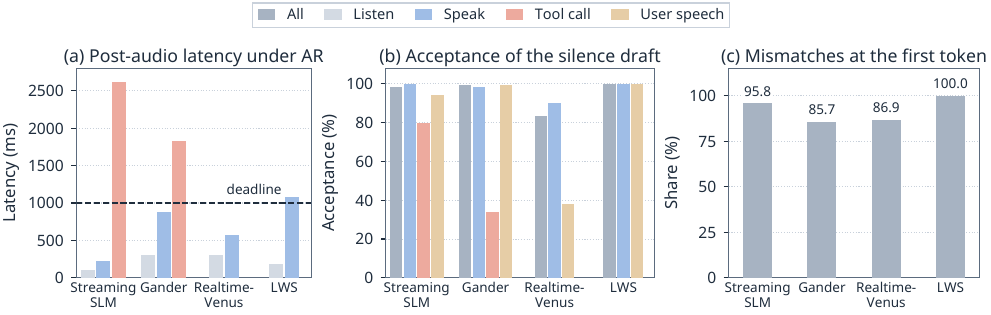}
  \caption{\textbf{Chunk profile and oracle probe.} (a) Mean post-audio latency by chunk type under AR; dashed line: 1\,s deadline. (b) Chunk-level acceptance rate of the silence-hypothesis draft over all chunks, speak chunks, tool-call chunks, and chunks with user speech. (c) Share of mismatched chunks that diverge at the first output token. Tool-call chunks and the bars of Streaming SLM, Gander, and Realtime-Venus in (a) use FDB~v3; all other bars use Llama Questions. Realtime-Venus and LWS produce no tool-call chunks.}
  \label{fig:observations}
\end{figure*}

\paragraph{Chunk profile.}
\label{sec:profile}
We profiled autoregressive decoding by chunk type on the four models of \cref{sec:setup} (\cref{fig:observations}a). Listen chunks process the audio and output only control tokens, and compute waits for the next audio during 70--89\% of the chunk duration. We call this time the \emph{idle window}. Speak and tool-call chunks sit at the other end. A tool call takes tens of serial forward passes and completes 1.8--2.6\,s after its audio arrives, and every evaluated tool call missed the deadline. LWS, whose speak chunks can also decode a written channel, missed the deadline on 95\% of them.

\subsection{Observations}
\label{sec:observations}
\label{sec:acceptance}
Using the idle window requires decoding chunk $t$ before $\mathbf{a}_t$ arrives, that is, under an assumed input. We take silence as the simplest assumption: the \emph{silence hypothesis} replaces $\mathbf{a}_t$ with $\mathbf{a}_{\text{sil}}$, the input tokens of a chunk of silence with no text input. To measure how often it yields the real output, we ran an oracle probe without a time limit. It decodes every chunk twice, once with $\mathbf{a}_t$ and once with $\mathbf{a}_{\text{sil}}$, and checks whether the two outputs are identical (\cref{fig:observations}b,c).

\paragraph{Observation 1: the next audio chunk usually does not change the output.}
Pooled over the four models, the two outputs match in 95.3\% of chunks. The match rate is lowest on Realtime-Venus while the user is speaking, when incoming speech updates the context (\cref{fig:observations}b and \cref{app:acceptance}).

\paragraph{Observation 2: changes concentrate at the first token.}
Pooled over the four models, 87.7\% of mismatched chunks diverge at the first output token, which in most models is the decision to keep listening or to start an output (\cref{fig:observations}c).

\subsection{Speculative Listening}
\label{sec:sl}
Speculative decoding~\citep{leviathan2023,chen2023} drafts the next few tokens and feeds the whole draft to the target model as input in one forward pass. In this forward pass, the distribution for the $i$-th token depends only on the first $i-1$ draft tokens. When these tokens are correct, it equals the distribution of autoregressive decoding at step $i$. Verification accepts the draft tokens up to the first mismatch and replaces the mismatched token with the choice of the target model. A single forward pass can then commit several tokens, and it takes about as long as decoding one token (\cref{sec:prelim}). \method{} uses the target model itself as the draft model and drafts chunk $t$ under the silence hypothesis during the idle window, before $\mathbf{a}_t$ arrives. It verifies the draft in the forward pass prefilling $\mathbf{a}_t$ (\cref{fig:method}). The forward passes of drafting run before the audio arrives. When the draft matches the whole output, the number of post-audio forward passes no longer grows with $|\mathbf{y}_t|$.

\paragraph{Drafting under the silence hypothesis.}
After chunk $t-1$ finishes, the model decodes greedily on its own KV cache,
\begin{equation}
\tilde{y}_i = \arg\max_{y \in \mathcal{V}} \, p_\theta\left(y \mid \mathbf{x}_{<t}, \mathbf{a}_{\text{sil}}, \tilde{\mathbf{y}}_{<i}\right), \quad i = 1, \dots, \gamma,
\label{eq:draft}
\end{equation}
until it reaches the end of the chunk or the idle window ends, which yields the primary draft $\tilde{\mathbf{y}} = (\tilde{y}_1, \dots, \tilde{y}_\gamma)$. When $\mathbf{a}_t$ arrives, the KV cache is rolled back to the end of $\mathbf{x}_{<t}$, and one forward pass over $(\mathbf{a}_t, \tilde{\mathbf{y}})$ prefills the audio and computes the target distributions $p_i = p_\theta(\cdot \mid \mathbf{x}_{<t}, \mathbf{a}_t, \tilde{\mathbf{y}}_{<i})$ for $i = 1, \dots, \gamma + 1$. Following greedy verification~\citep{specbench}, the number of accepted tokens is
\begin{equation}
n = \min\left(\left\{\, i - 1 \mid 1 \le i \le \gamma,\ \tilde{y}_i \neq \arg\max p_i \,\right\} \cup \{\gamma\}\right).
\label{eq:accept}
\end{equation}
The forward pass commits $\tilde{y}_1, \dots, \tilde{y}_n$ and the correction $\arg\max p_{n+1}$, and decoding continues from there. The committed output is identical to \cref{eq:ar}. Every forward pass commits at least one token, and the number of post-audio forward passes never exceeds the number under autoregressive decoding. Under stochastic decoding, the draft is verified by speculative sampling~\citep{leviathan2023,chen2023}, which preserves the target distribution (\cref{sec:robust}).

\paragraph{Hypothesis pool and runner-up branch.}
The idle window of one chunk holds a short speak chunk but may end before a tool call of several dozen tokens is drafted. A user request spans several chunks, and drafts made during earlier chunks may already contain part of the eventual output. \method{} keeps all drafts of a session in a \emph{hypothesis pool} $\mathbf{C}$, indexed by their $N$-grams as in lookahead decoding~\citep{lookahead}. Observation~2 shows that most mismatches start at the first token. When the primary draft finishes before the idle window ends, the model decodes a second draft, the \emph{runner-up branch}, from the runner-up first token $\arg\max_{y \in \mathcal{V} \setminus \{\tilde{y}_1\}} p_\theta(y \mid \mathbf{x}_{<t}, \mathbf{a}_{\text{sil}})$, and adds it to $\mathbf{C}$. During drafting and verification, after each forward pass the runtime looks up the longest $N$-gram in $\mathbf{C}$ and in the session's earlier outputs matching the suffix of the output so far, and appends its continuation to the following forward pass. The number of post-audio forward passes then grows only with output tokens missing from every draft and with the appended pool segments (\cref{sec:length}).

\section{TickTock-Bench}
\label{sec:bench}
\begin{table}[t]
\caption{\textbf{Composition of TickTock-Bench.} \#Sess.: number of sessions. \#Chunks: number of 1\,s chunks, fixed by the input audio.}
\label{tab:bench}
\vskip 0.1in
\centering
\small
\setlength{\tabcolsep}{2.8pt}
\begin{tabular}{@{}lllrr@{}}
\toprule
\textbf{Category} & \textbf{Task} & \textbf{Focus} & \textbf{\#Sess.} & \textbf{\#Chunks} \\
\midrule
\multirow{5}{*}{\shortstack[l]{Speech\\reasoning}}
  & Web Q.     & factual QA& 100 & 1,345 \\
  & Llama Q.   & factual QA& 100 & 1,423 \\
  & TriviaQA   & factual QA& 100 & 1,671 \\
  & AlpacaEval & open-ended answers& 100 & 1,598 \\
  & BBA        & multi-step reasoning    & 60  & 1,825 \\
\midrule
\multirow{2}{*}{\shortstack[l]{Conv.\\dynamics}}
  & FDB v1     & turn-taking behaviors & 100 & 1,885 \\
  & FDB v1.5   & overlapping speech & 100 & 1,371 \\
\midrule
Agentic & FDB v3 & tool calls & 100 & 4,765 \\
\midrule
Total & & & 760 & 15,883 \\
\bottomrule
\end{tabular}
\end{table}

In this section, we introduce TickTock-Bench, a benchmark for evaluating decoding methods on chunk-interleaved full-duplex models. In these models, the chunk duration acts as a hard deadline: a chunk that takes longer to decode delays the following chunks, and the conversation lags behind real time. The model then processes the user's speech late, and its responses become stale in the ongoing interaction. The value of a decoding method for these models lies in how it changes this behaviour: how long a response or a tool call waits after the user's audio, how often chunks miss their deadlines, and how far the conversation falls behind. Measuring these effects requires checking each chunk against its deadline, by chunk type, in the model's own runtime. Chunk-interleaved models are also growing in number, each with its own runtime, and a shared benchmark lets decoding methods be compared across them. Following Spec-Bench~\citep{specbench}, which compared speculative decoding methods on text generation under a unified protocol, TickTock-Bench evaluates all methods on the same sessions and hardware under this chunk-level measurement. We release it for evaluating future decoding methods on these models in the same way.

TickTock-Bench covers three task categories (\cref{tab:bench}): speech reasoning with Web Questions~\citep{webq}, Llama Questions~\citep{llamaq}, and TriviaQA~\citep{triviaqa} from OpenAudioBench~\citep{openaudiobench}, AlpacaEval from VoiceBench~\citep{voicebench}, and Big Bench Audio (BBA)~\citep{bba,bigbench,bbh}; conversational dynamics with Full-Duplex-Bench (FDB) v1~\citep{fdb} and v1.5~\citep{fdb15}; and agentic tool use with FDB~v3~\citep{fdb3}. Speech reasoning tests spoken question answering, conversational dynamics tests listening and responding around pauses, turn ends, and overlapping speech, and agentic tool use tests tool calls.

The replay harness feeds each session to the model chunk by chunk through its own runtime, one session at a time on one NVIDIA A100 GPU, and labels each chunk by its output as a listen, speak, or tool-call chunk. We use the speedup over autoregressive decoding in mean post-audio latency per chunk type as the primary metric, measured in wall-clock time without speech synthesis. Computation that a method performs before the audio arrives runs in the idle window, and any part beyond the idle window is added to the post-audio latency. To measure how far decoding falls behind real time, we also report the deadline miss rate and the maximum lag behind the audio per session, together with \#MAT~\citep{specbench}. Details are provided in \cref{app:bench}.

\section{Experiments}
\label{sec:exp}

\begin{table*}[t]
\caption{\textbf{Main results on TickTock-Bench.} Speedup ($\times$) over autoregressive decoding (AR) in mean post-audio latency of speak chunks across four chunk-interleaved full-duplex models (1\,s chunk duration, greedy decoding, batch size 1). $^{\dagger}$FDB~v3 reports tool-call chunks for Streaming SLM and Gander, and speak chunks otherwise. \#MAT denotes the number of output tokens per target forward pass over all chunks relative to AR ($= 1.00$), averaged over tasks (per-task values in \cref{app:mat}); Avg.\ is the mean speedup over the seven speech-reasoning and conversational-dynamics tasks.}
\label{tab:main}
\vskip 0.1in
\centering
\small
\setlength{\tabcolsep}{2.4pt}
\renewcommand{\arraystretch}{1.05}
\begin{tabular}{@{}ll ccccc cc c c c@{}}
\toprule
\multirow{2}{*}{\textbf{Model}} & \multirow{2}{*}{\textbf{Method}} & \multicolumn{5}{c}{\textbf{Speech Reasoning}} & \multicolumn{2}{c}{\textbf{Conv.\ Dynamics}} & \textbf{Agentic} & \multirow{2}{*}{\textbf{\#MAT}} & \multirow{2}{*}{\textbf{Avg.}} \\
\cmidrule(lr){3-7}\cmidrule(lr){8-9}\cmidrule(lr){10-10}
& & \textbf{Web Q.} & \textbf{Llama Q.} & \textbf{TriviaQA} & \textbf{AlpacaEval} & \textbf{BBA} & \textbf{FDB v1} & \textbf{FDB v1.5} & \textbf{FDB v3}$^{\dagger}$ & & \\
\midrule
\multirow{4}{*}{\shortstack[l]{Streaming\\SLM}}
  & PLD              & 1.01 & 1.01 & 1.02 & 1.01 & 1.03 & 1.03 & 1.02 & 1.01 & 1.01 & 1.02 \\
  & Token Recycling  & 1.01 & 1.00 & 1.02 & 1.01 & 1.03 & 1.03 & 1.02 & 1.02 & 1.01 & 1.02 \\
  & Lookahead        & 0.98 & 0.98 & 0.99 & 0.97 & 0.99 & 0.98 & 1.00 & 1.07 & 1.03 & 0.99 \\
  & \ours           & 2.80 & 2.87 & 3.04 & 2.75 & 2.67 & 2.62 & 2.57 & 23.41 & 2.63 & 2.76 \\
\midrule
\multirow{4}{*}{Gander}
  & PLD              & 1.04 & 1.02 & 1.01 & 1.07 & 1.03 & 1.07 & 1.06 & 1.07 & 1.21 & 1.04 \\
  & Token Recycling  & 1.01 & 0.99 & 1.00 & 1.02 & 0.98 & 1.01 & 1.02 & 1.08 & 1.01 & 1.00 \\
  & Lookahead        & 1.00 & 0.97 & 0.98 & 1.00 & 0.96 & 1.01 & 1.00 & 1.12 & 1.01 & 0.99 \\
  & \ours           & 2.82 & 3.00 & 3.26 & 2.18 & 2.47 & 2.18 & 2.27 & 3.45 & 2.20 & 2.60 \\
\midrule
\multirow{4}{*}{\shortstack[l]{Realtime-\\Venus}}
  & PLD              & 1.11 & 1.08 & 1.10 & 1.12 & 1.11 & 1.11 & 1.11 & 1.05 & 1.23 & 1.11 \\
  & Token Recycling  & 1.03 & 1.01 & 1.02 & 1.01 & 1.03 & 1.01 & 1.01 & 0.99 & 1.01 & 1.02 \\
  & Lookahead        & 1.00 & 0.99 & 1.00 & 1.01 & 0.99 & 0.99 & 0.99 & 0.96 & 1.01 & 1.00 \\
  & \ours           & 2.75 & 2.81 & 2.73 & 2.85 & 2.59 & 2.57 & 2.54 & 2.41 & 2.15 & 2.69 \\
\midrule
\multirow{4}{*}{LWS}
  & PLD              & 1.09 & 1.10 & 1.11 & 1.09 & 1.17 & 1.12 & 1.09 & 1.14 & 1.11 & 1.11 \\
  & Token Recycling  & 1.14 & 1.13 & 1.15 & 1.12 & 1.19 & 1.17 & 1.13 & 1.17 & 1.13 & 1.15 \\
  & Lookahead        & 1.14 & 1.13 & 1.14 & 1.14 & 1.23 & 1.14 & 1.12 & 1.15 & 1.15 & 1.15 \\
  & \ours           & 2.43 & 2.95 & 2.74 & 2.24 & 3.46 & 2.37 & 2.55 & 1.60 & 2.52 & 2.68 \\
\bottomrule
\end{tabular}
\end{table*}

\subsection{Experimental Setup}
\label{sec:setup}
\paragraph{Evaluation.}
On FDB~v3, we reported tool-call chunks for Streaming SLM and Gander, and speak chunks for Realtime-Venus and LWS. Realtime-Venus issued no tool calls on these sessions under greedy decoding, and LWS has no tool-call channel. The ablation and the analysis use FDB~v3 and Llama Questions. All evaluated methods are lossless: in exact arithmetic, their outputs are identical to those of autoregressive decoding (\cref{sec:quality}).

\paragraph{Models.}
We evaluated four chunk-interleaved full-duplex models with a 1\,s chunk duration, covering more than one model family:
(1) \emph{Streaming SLM}, a chunk-interleaved full-duplex model that we trained from the turn-based Qwen2.5-Omni-7B Thinker, with audio input, text input, and tool calls;
(2) \emph{Gander}~\citep{gander} and (3) \emph{Realtime-Venus}~\citep{venus}, two public agentic full-duplex models built on MiniCPM-o~\citep{minicpmo45}, run with their released runtimes; and
(4) \emph{LWS}~\citep{lws}, a tri-channel full-duplex model built on MiniCPM-o whose speak chunks can decode a written text channel alongside the text it speaks. In all four models, the backbone (Thinker) decodes text tokens, and a separate speech decoder (Talker) synthesizes the audio from these tokens or their hidden states. We measured the Thinker only, since it is the only component in which the methods differ.

\paragraph{Baselines.}
We compared \method{} with autoregressive decoding (AR) and three training-free, model-free speculative decoding methods that run after the audio arrives. Prompt Lookup Decoding (PLD)~\citep{pld} drafts tokens by matching $n$-grams in the context. Token Recycling~\citep{tokenrecycling} drafts from candidate tokens recycled from earlier decoding steps, and lookahead decoding~\citep{lookahead} drafts from $n$-grams collected by Jacobi iterations; we evaluated both in chain verification mode to match the models' runtimes.

\paragraph{Implementation.}
All models ran in bfloat16 with greedy decoding and batch size 1, under the protocol of \cref{sec:bench}.

\begin{table}[t]
\caption{\textbf{Maximum lag behind the incoming audio per session}. Mean and Max: mean and maximum over sessions (ms). Lagging: share of sessions whose maximum lag exceeds 1\,s (\%). FDB~v3.}
\label{tab:backlog}
\vskip 0.1in
\centering
\small
\setlength{\tabcolsep}{2.5pt}
\begin{tabular}{@{}l rrr rrr@{}}
\toprule
\multirow{2}{*}{\textbf{Model}} & \multicolumn{3}{c}{\textbf{AR}} & \multicolumn{3}{c}{\textbf{\method{}}} \\
\cmidrule(lr){2-4}\cmidrule(lr){5-7}
& \textbf{Mean} & \textbf{Max} & \textbf{Lagging} & \textbf{Mean} & \textbf{Max} & \textbf{Lagging} \\
\midrule
Streaming SLM  & 568 & 2619 & 35 & 0  & 0    & 0 \\
Gander         & 444 & 1227 & 9  & 42 & 675  & 0 \\
Realtime-Venus & 1   & 108  & 0  & 0  & 0    & 0 \\
LWS            & 430 & 3966 & 20 & 55 & 2168 & 1 \\
\bottomrule
\end{tabular}
\end{table}

\begin{figure*}[t]
  \centering
  \includegraphics[width=\textwidth]{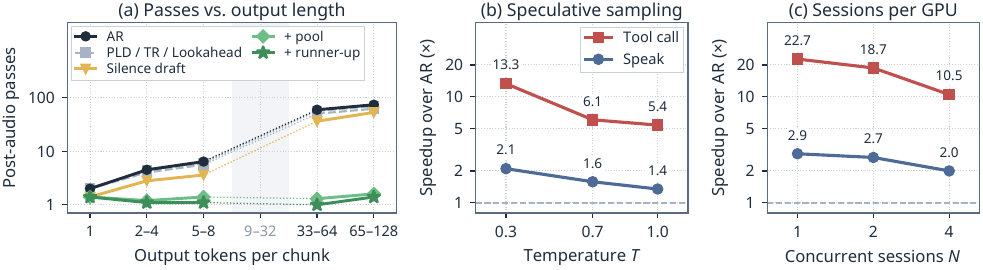}
  \caption{\textbf{In-depth analysis on Streaming SLM} (FDB~v3). (a) Post-audio forward passes per chunk by output length; the shaded bin (9--32 tokens) contains no chunks. (b) Speedup over AR under speculative sampling at temperature $T$. (c) Speedup over AR with $N$ concurrent sessions on one A100 GPU, in wall-clock time.}
  \label{fig:analysis}
\end{figure*}

\subsection{Main Results}
\label{sec:main}
\paragraph{Speedup across models and tasks.}
We examined whether decoding in the idle window shortens post-audio latency across models and tasks. As shown in \cref{tab:main}, we observe two main trends. First, \method{} is faster than autoregressive decoding and the three training-free baselines on all models and tasks. It reduces mean speak-chunk latency by 2.6--2.8\x{} on average across the four models and tool-call latency on FDB~v3 by 23.41\x{} on Streaming SLM (\cref{fig:case_long}b) and 3.45\x{} on Gander; the three baselines reach 0.96--1.23\x{}. The number of tokens committed per forward pass shows the same gap: \method{} reaches a \#MAT of 2.15--2.63, and the baselines reach 1.01--1.23. Second, on Streaming SLM, the speedup grows with output length: speak chunks contain a few tokens and are sped up by 2.6--3.0\x{}, and tool calls contain several dozen tokens and are sped up by 23.41\x{}. Both trends follow from the source of the drafts and the way they are verified. \method{} drafts with the target model before the audio arrives, and as Observation~1 shows, most drafts match the real output. The baselines draft from $n$-grams in the context after the audio arrives, and these $n$-grams rarely match the new response in a speak chunk. An accepted draft is verified within the audio forward pass and saves one forward pass per output token under autoregressive decoding, and longer outputs save more. By drafting in the idle window and verifying in the audio forward pass, \method{} decodes most of a chunk's output before its audio arrives.

\paragraph{Deadline misses.}
We further examined whether \method{} keeps the models in real time. A deadline miss delays the following chunks, and the delay accumulates over consecutive chunks. To quantify this effect, we measured the maximum lag behind the audio per session (\cref{app:bench}). As shown in \cref{tab:backlog}, under AR the maximum lag reaches 2.6\,s on Streaming SLM and 4.0\,s on LWS, and 35\% of the sessions of Streaming SLM lag behind the audio by more than 1\,s. With \method{}, the lag stays at zero on Streaming SLM and Realtime-Venus and averages 42\,ms on Gander and 55\,ms on LWS, and at most 1\% of the sessions lag behind the audio by more than 1\,s. These results show that \method{} reduces the maximum lag behind real time by 94.4\% on average.

\begin{table}[t]
\caption{\textbf{Ablation study.} Speedup ($\times$) over AR as the components of \method{} are added. Speak: Llama Questions; tool: FDB~v3. Best value per column in bold.}
\label{tab:ablation}
\vskip 0.1in
\centering
\small
\setlength{\tabcolsep}{4pt}
\begin{tabular}{@{}l cc cc@{}}
\toprule
& \multicolumn{2}{c}{\textbf{Streaming SLM}} & \multicolumn{2}{c}{\textbf{Gander}} \\
\cmidrule(lr){2-3}\cmidrule(lr){4-5}
\textbf{Method} & \textbf{Speak} & \textbf{Tool} & \textbf{Speak} & \textbf{Tool} \\
\midrule
AR & 1.00 & 1.00 & 1.00 & 1.00 \\
+ silence-hypothesis draft & \textbf{2.88} & 1.42 & 2.64 & 1.21 \\
+ hypothesis pool & 2.87 & 22.04 & 2.97 & 2.04 \\
+ runner-up branch & 2.87 & \textbf{23.41} & \textbf{3.00} & \textbf{3.45} \\
\bottomrule
\end{tabular}
\end{table}

\begin{figure*}[t]
  \centering
  \includegraphics[width=\textwidth]{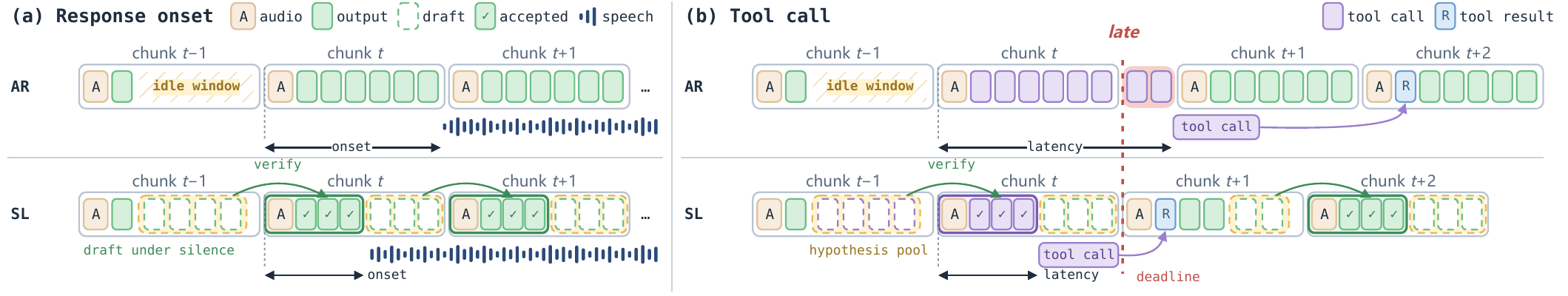}
  \caption{\textbf{Case study of user-perceived latency.} SL: \method{}. (a) Response onset. Under AR, speech starts after the last token of the chunk is decoded. With \method{}, the draft made under silence is verified when the audio arrives, and speech starts earlier. (b) Tool call. Under AR, a long tool call runs past the deadline and delays the next chunk. With \method{}, the call drafted in earlier chunks is verified from the hypothesis pool and is ready within the chunk.}
  \label{fig:case}
\end{figure*}

\subsection{Ablation Study}
\label{sec:ablation}
To examine the contribution of each component, we added silence-hypothesis drafting, the hypothesis pool, and the runner-up branch to AR one at a time on Streaming SLM and Gander, using FDB~v3 for tool-call chunks and Llama Questions for speak chunks.

\paragraph{Effects of silence-hypothesis drafting.}
We first added silence-hypothesis drafting alone. As shown in \cref{tab:ablation}, it speeds up speak chunks by 2.6--2.9\x{} and tool-call chunks by 1.2--1.4\x{}. These results show that drafting within one idle window covers short outputs. A tool call of several dozen tokens exceeds the idle window, and its draft stops before the call is complete.

\paragraph{Effects of the hypothesis pool.}
We then added the hypothesis pool. As shown in \cref{tab:ablation}, it raises the tool-call speedup from 1.42\x{} to 22.04\x{} on Streaming SLM and from 1.21\x{} to 2.04\x{} on Gander. These results show that drafts from earlier chunks cover most of a tool call: a user request spans several chunks, and the drafts made in earlier chunks already contain most of the eventual call.

\paragraph{Effects of the runner-up branch.}
Finally, we added the runner-up branch. As shown in \cref{tab:ablation}, it raises the tool-call speedup from 2.04\x{} to 3.45\x{} on Gander and from 22.04\x{} to 23.41\x{} on Streaming SLM. These results show that the branch helps where the silence draft departs at the first token (Observation~2): it provides a draft starting from the alternative first token. On Streaming SLM, the pool already leaves little for the branch to remove.

\begin{table}[t]
\caption{\textbf{User-perceived latency} (ms). Onset: response onset latency, averaged over the eight tasks. Tool: tool-call latency, on FDB~v3. Best value per model in bold.}
\label{tab:ux}
\vskip 0.1in
\centering
\small
\setlength{\tabcolsep}{4pt}
\begin{tabular}{@{}l rr rr@{}}
\toprule
\multirow{2}{*}{\textbf{Model}} & \multicolumn{2}{c}{\textbf{AR}} & \multicolumn{2}{c}{\textbf{\method{}}} \\
\cmidrule(lr){2-3}\cmidrule(lr){4-5}
& \textbf{Onset} & \textbf{Tool} & \textbf{Onset} & \textbf{Tool} \\
\midrule
Streaming SLM  & 417  & 2623 & \best{74}  & \best{112} \\
Gander         & 763  & 1827 & \best{228} & \best{513} \\
Realtime-Venus & 565  & \na  & \best{241} & \na \\
LWS            & 1089 & \na  & \best{204} & \na \\
\bottomrule
\end{tabular}
\end{table}

\subsection{In-Depth Analysis}
\label{sec:in_depth}
\paragraph{Analysis on user-perceived latency.}
\label{sec:case}
To examine whether the speedup shortens the delays that users hear, we measured two latencies (\cref{fig:case} and \cref{app:case}). The response onset latency is the time from the arrival of the audio of the first speak chunk of a response to the end of its Thinker output. The tool-call latency is the time from the arrival of the audio of a tool-call chunk to the end of the call. As shown in \cref{tab:ux}, \method{} reduces the response onset latency by 73\% on average across the four models and the tool-call latency by 72--96\%. These results show that the post-audio speedup shortens the delays before a response and before a tool call. Human speakers start a response about 200\,ms after the other speaker ends~\citep{levinson2015}. Under AR, decoding alone adds 417--1089\,ms before a response starts. With \method{}, it adds 74--241\,ms, close to this gap, and tool calls become ready within the chunk duration on average.

\paragraph{Analysis on output length and sampling.}
\label{sec:length}\label{sec:robust}
To examine whether the speedup holds beyond the main setting, we varied the output length and the sampling temperature $T \in \{0.3, 0.7, 1.0\}$ (\cref{fig:analysis}a,b). With \method{}, a chunk needs 1.0--1.4 post-audio forward passes for outputs of 2 to 128 tokens. AR grows from 2.0 passes for one token to 73.6 passes for 65--128 tokens, the three baselines follow the same trend, and silence-hypothesis drafting alone reaches 52.9 passes. \method{} speeds up speak and tool-call chunks by 2.1\x{} and 13.3\x{} at $T = 0.3$ and by 1.4\x{} and 5.4\x{} at $T = 1.0$. Sampled tokens depart from the greedy draft more often as the temperature rises and shorten the accepted prefix. These results show that \method{} needs few post-audio passes at every output length and remains effective under sampling.

\paragraph{Analysis on response quality.}
\label{sec:quality}
To examine whether \method{} preserves the outputs of AR, we compared the two. Under float32, \method{} reproduces AR token for token on nearly all sessions. Under bfloat16, used in the main tables, rounding can change outputs, and the AlpacaEval judge scores of AR and \method{} differ by less than 0.1 point on all four models (\cref{tab:quality} in \cref{app:quality}). These results show that \method{} preserves the response quality of AR.

\section{Conclusion and Future Work}
\label{sec:conclusion}
In this work, we studied decoding latency in chunk-interleaved full-duplex speech language models. We identified an asymmetric workload across chunks: listen chunks leave compute idle while awaiting audio, and speak and tool-call chunks require serial generation growing with output length. Comparing outputs under real audio and the silence hypothesis, we found two properties: the next audio chunk usually does not change the output, and changes concentrate at the first token. Building on these properties, we proposed \method{}, a training-free and lossless framework. It drafts each chunk in the idle window and verifies the draft in the audio forward pass. On TickTock-Bench, it reduces the maximum lag behind real time by 94.4\% and the decoding delay before a response starts by 73\% on average. Future work includes drafting under input hypotheses beyond digital silence, such as streaming acoustic features.

\section*{Limitations}
This work has three limitations. First, \method{} targets low-concurrency, latency-sensitive deployment. It computes during the idle window, and the idle window shrinks as more sessions share a GPU or are batched. In our multi-session experiment, concurrent sessions share one GPU, and the speedup decreases as the number of sessions grows (\cref{app:concurrency}). Second, we compared the Thinker only, since the Talkers of the four models synthesize speech in different ways and would confound the comparison. Third, our formulation targets chunk-interleaved architectures with an idle window between chunks; parallel-stream architectures such as Moshi~\citep{moshi}, which decode at every 80\,ms audio frame, require a different formulation.

\bibliography{refs}
\bibliographystyle{icml2026}

\newpage
\appendix
\onecolumn
\section{Benchmark Details}
\label{app:bench}

We release the session manifests, the replay harness, and the evaluation scripts. Latencies measured on other hardware or under other CPU load differ from ours. The lag accumulates over chunks as $b_t = \max(0,\, b_{t-1} + \ell_t - \text{chunk duration})$ with $b_0 = 0$, where $\ell_t$ is the post-audio latency of chunk $t$. The response onset and tool-call latencies in \cref{sec:in_depth} include this lag. Drafting for chunk $k$ starts when the computation of chunk $k-1$ ends, and its budget is the chunk duration minus the measured latency of chunk $k-1$, which includes the audio encoder and runtime overheads. Drafting beyond the budget is added to the post-audio latency of chunk $k$. Under AR and the baselines, the GPU stays idle during the idle window.

\section{Additional Analyses}
\label{app:analyses}

\subsection{Response Quality}
\label{app:quality}
We scored the spoken responses of AR and \method{} on the 100 AlpacaEval sessions with the open-ended rubric of VoiceBench~\citep{voicebench,flexislm} (scores 1--5), using DeepSeek-V4.1-Flash as the judge.

\begin{table}[h]
\caption{\textbf{Response quality on AlpacaEval.} Judge score (1--5) of AR and \method{}, and their difference $\Delta$.}
\label{tab:quality}
\vskip 0.1in
\centering
\small
\begin{tabular}{@{}l ccc@{}}
\toprule
\textbf{Model} & \textbf{AR} & \textbf{\method{}} & $\boldsymbol{\Delta}$ \\
\midrule
Streaming SLM  & 3.10 & 3.13 & $+$0.03 \\
Gander         & 3.56 & 3.63 & $+$0.07 \\
Realtime-Venus & 2.98 & 3.00 & $+$0.02 \\
LWS            & 3.02 & 3.07 & $+$0.05 \\
\bottomrule
\end{tabular}
\end{table}

\subsection{Acceptance by Chunk Type}
\label{app:acceptance}
\cref{tab:acceptance} lists the chunk-level acceptance rates of the oracle probe (\cref{sec:acceptance}) by chunk type.

\begin{table}[h]
\caption{\textbf{Chunk-level acceptance rate (\%) of the silence-hypothesis draft} by chunk type. First token: share (\%) of mismatched chunks that diverge at the first output token. Llama Questions, 1,423 chunks per model; tool-call chunks on FDB~v3. LWS has a single mismatched chunk.}
\label{tab:acceptance}
\vskip 0.1in
\centering
\small
\begin{tabular}{@{}l cccccc@{}}
\toprule
\textbf{Model} & \textbf{All} & \textbf{Listen} & \textbf{Speak} & \textbf{Tool call} & \textbf{User speech} & \textbf{First token} \\
\midrule
Streaming SLM  & 98.3 & 97.6 & 99.8  & 80.0 & 94.2  & 95.8 \\
Gander         & 99.5 & 99.6 & 98.6  & 34.0 & 99.5  & 85.7 \\
Realtime-Venus & 83.3 & 81.4 & 90.0  & \na  & 37.9  & 86.9 \\
LWS            & 99.9 & 99.8 & 100.0 & \na  & 100.0 & 100.0 \\
\bottomrule
\end{tabular}
\end{table}

\subsection{Per-Benchmark \#MAT}
\label{app:mat}
\begin{table}[H]
\caption{\#MAT of \method{} across benchmarks (output tokens per target forward pass over all chunks of a task, normalized to 1.00 for AR).}
\label{tab:mat}
\vskip 0.1in
\centering
\small
\begin{tabular}{@{}l cccccccc c@{}}
\toprule
\textbf{Model} & \textbf{Web Q.} & \textbf{Llama Q.} & \textbf{TriviaQA} & \textbf{AlpacaEval} & \textbf{BBA} & \textbf{FDB v1} & \textbf{FDB v1.5} & \textbf{FDB v3} & \textbf{Avg.} \\
\midrule
Streaming SLM  & 3.15 & 2.87 & 2.77 & 3.28 & 2.03 & 2.11 & 3.06 & 1.78 & 2.63 \\
Gander         & 2.33 & 2.22 & 2.18 & 2.40 & 2.01 & 2.10 & 2.44 & 1.96 & 2.20 \\
Realtime-Venus & 2.40 & 2.31 & 2.21 & 2.31 & 1.73 & 1.95 & 2.45 & 1.82 & 2.15 \\
LWS            & 2.63 & 3.16 & 2.83 & 2.40 & 2.82 & 2.16 & 2.62 & 1.58 & 2.52 \\
\bottomrule
\end{tabular}
\end{table}

\subsection{Multi-Session Serving}
\label{app:concurrency}
To examine \method{} under GPU contention, we ran $N \in \{1, 2, 4\}$ concurrent sessions on one A100 GPU and measured wall-clock latency. As shown in \cref{fig:analysis}c, at $N = 1$ \method{} speeds up speak chunks by 2.9\x{} and tool-call chunks by 22.7\x{}. At $N = 4$, it speeds up speak chunks by 2.0\x{} and tool-call chunks by 10.5\x{}, and it reduces the deadline miss rate of tool-call chunks from 100\% to 2.9\%. These results show that \method{} remains effective under GPU contention, and its gain on long tool calls decreases as $N$ grows. Concurrent sessions slow down drafting, which leaves less time to draft long tool calls.

\section{Case Study}
\label{app:case}
\begin{figure}[h]
  \centering
  \includegraphics[width=\textwidth]{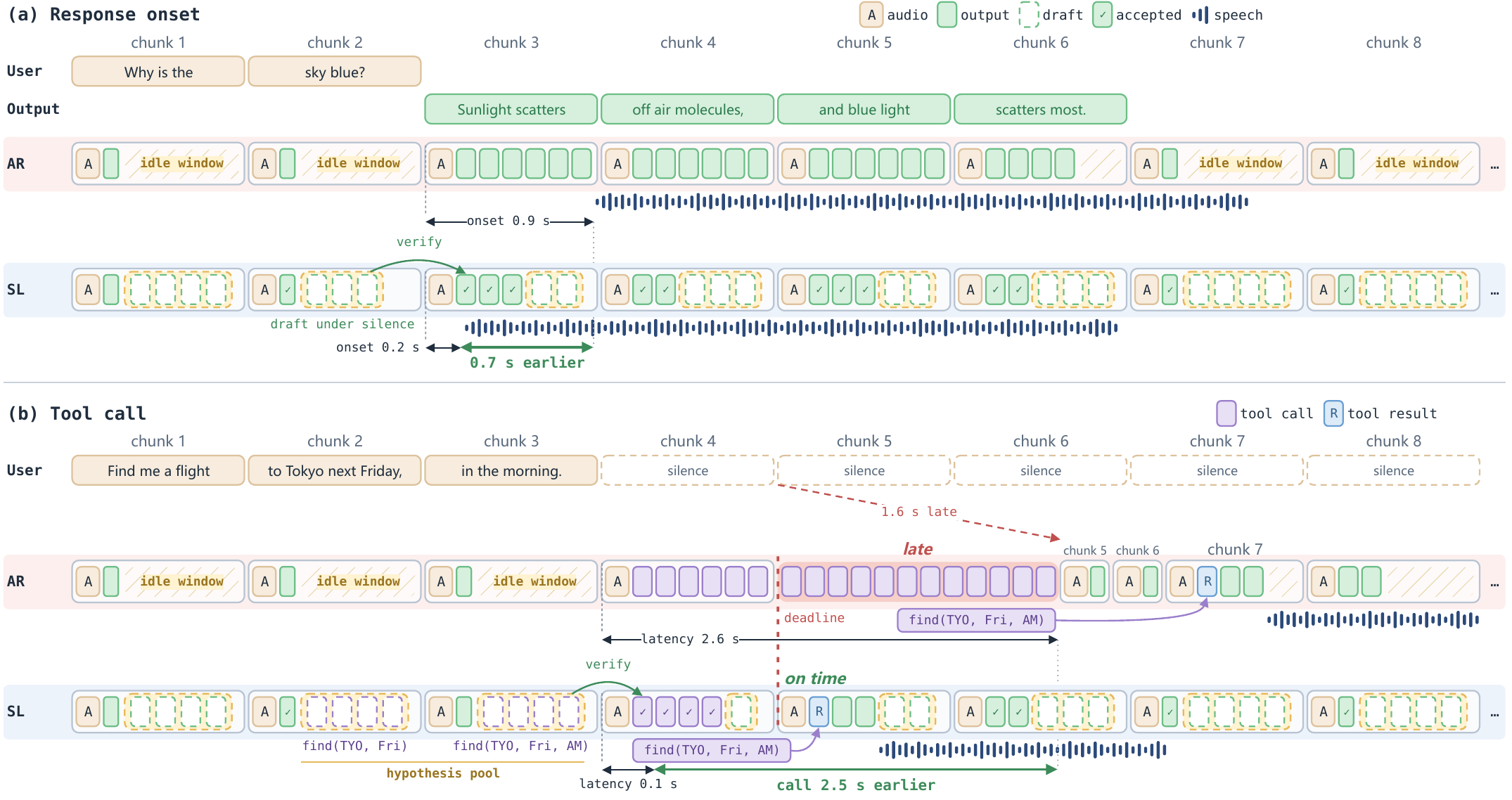}
  \caption{\textbf{Case study on two illustrative sessions.} SL: \method{}. Span labels give typical measured latencies (\cref{tab:ux}), and green spans mark the time saved by \method{}. Check marks denote drafted tokens accepted by verification. (a) Response onset. The first speak chunk is drafted under silence in chunk~2 and verified when its audio arrives. (b) Tool call. R: tool result. Drafts made while the user speaks contain the tool call and enter the hypothesis pool. Under AR, the call misses the deadline, and the dashed arrow links the audio of chunk~5 to the time AR starts processing it.}
  \label{fig:case_long}
\end{figure}

To illustrate how \method{} shortens the delays that users perceive~\citep{aural,thinkaloud}, we walk through two example sessions in \cref{fig:case_long}. The latencies in the figure are typical values measured on TickTock-Bench (\cref{tab:ux}).

\paragraph{Response onset.}
As shown in \cref{fig:case_long}a, the user asks a question in chunks 1 and 2, and the model starts its answer in chunk 3. Under AR, the model decodes the first speak chunk after its audio arrives, and the output is ready after 0.9\,s. With \method{}, the model drafts chunk 3 under silence during the idle window of chunk 2, and the audio forward pass of chunk 3 verifies the draft. The output is ready after 0.2\,s. In chunks 4 to 6, both methods finish each speak chunk within the chunk duration, and each chunk starts when its audio arrives. The answer then streams at the same pace under both methods and starts 0.7\,s earlier with \method{}.

\paragraph{Tool call.}
As shown in \cref{fig:case_long}b, the user requests a flight over chunks 1 to 3. During each idle window, \method{} drafts the next chunk under silence. From chunk 2 on, these drafts contain the tool call with the details heard so far, and the hypothesis pool keeps them. While the user is still speaking, verification rejects them at the first token. In chunk 4, the audio forward pass verifies the drafted call, and the call is ready after 0.1\,s. Under AR, the model decodes the call token by token for 2.6\,s and misses the deadline by 1.6\,s. Chunk 5 then starts 1.6\,s after its audio arrives. In this example, the tool returns its result 0.5\,s after the call is dispatched, and the result enters the first chunk that starts after it returns~\citep{jarvis,duplexagentrsi}. It enters chunk 5 with \method{} and chunk 7 under AR. The answer to the user then starts 2.2\,s earlier with \method{}.

\end{document}